\documentclass[runningheads]{llncs}

\usepackage{header}
\usepackage{mathtools}

\begin{document}
\title{An HPC Approach to Accelerate\\ Tensor Decompositions}
%
%
\author{Markus Hellgren\inst{1} \and
Erna Begovi\'c~Kova\v{c}\inst{2}\orcidID{0000-0002-3213-1465} \and Hans O. Karlsson\inst{1}\orcidID{0000-0002-5366-4949} \and
Roman Iakymchuk\inst{1,3}\orcidID{0000-0003-2414-700X}}
\authorrunning{M.~Hellgren et al.}
%
\institute{Uppsala University, Department of Information Technology, Sweden\\
\email{markus.hellgren@gmail.com, hans.karlsson@it.uu.se, roman.iakymchuk@it.uu.se}
\and
University of Zagreb, Faculty of Chemical Engineering and Technology, Croatia\\
\email{ebegovic@fkit.unizg.hr}
\and
Ume\aa{} University, Department of Computing Science, Sweden}
\maketitle              
\begin{abstract}
Quantum systems grow in complexity so rapidly that even modest models become difficult to simulate, creating a strong need for methods that can handle high-dimensional data, also known as tensors. In this work, we investigate a novel Jacobi-type tensor algorithm for tensor decomposition and develop a CUDA-based algorithm that supports tensors of arbitrary order on a single GPU. We test the implementation on NVIDIA H100 GPUs and show that the algorithm converges correctly for diagonalizable tensors up to nine dimensions, with runtime scaling in a predictable way as tensor order grows. Finally, our general algorithm outperforms the original MATLAB reference by more than two orders of magnitude.

\end{abstract}
\section{Introduction}
The curse of dimensionality appears across many fields of science and engineering, and quantum physics is no exception. As the number of particles increases, the dimensionality of the Hilbert space grows exponentially, making even relatively small quantum systems difficult to treat. For example, an $N$-qubit quantum computer has a state space of dimension $2^N$, which increases exponentially with each additional qubit. Tensor decomposition techniques are essential for describing strongly correlated systems in condensed matter~\cite{RevModPhys}, quantum dynamics of high-dimensional molecular vibrations~\cite{Larsson24} as well as  quantum simulations of spin transport in one-dimensional Heisenberg chains~\cite{digitalquamtumspin26}.

This challenge is driving development and evaluation of new algorithms and implementations that can handle high‑dimensional problems more efficiently, particularly in high performance computing (HPC) environments where large‑sca\-le simulations are carried out. In the context of quantum physics, representative examples of such studies are presented in \cite{Menczer25} and \cite{Borrelli_2020}. In particular, there is strong motivation for tools that support spectral analysis, which plays a central role in understanding the properties of quantum systems. Many of the relevant models, operators, and data structures in this context can naturally be expressed as higher‑order tensors (multidimensional arrays), making tensor‑based methods appealing.

Tensor decompositions are surveyed in~\cite{kolda-review09}. Tucker methods such as HOSVD and HOOI~\cite{Lathauwer00,Lathauwer00-2} produce a generally dense core, whereas orthogonal tensor diagonalization maximizes the norm of its diagonal. Jacobi-type algorithms have been developed for symmetric and general third-order tensors~\cite{ULC183,ULC19,ULC20,begovic_kovac_convergence_2023,BegovicPerkovic24}. The present work extends the third-order approach of~\cite{begovic_kovac_convergence_2023} to arbitrary-order tensors and develops a GPU implementation based on parallel pivot sets. Although parallel and GPU-accelerated Jacobi methods are well established for matrix problems~\cite{LukPark89,DrmacVeselic07,Novakovic14}, comparable implementations for diagonalizing general dense tensors of arbitrary order appear to be unavailable.

Our main contributions:
\begin{itemize}
    \item based on the theoretical derivation with the corresponding Matlab implementation for the third order tensors~\cite{begovic_kovac_convergence_2023}, we designed a general algorithm for the Jacobi-type tensor decomposition, see~\Cref{sec:general}, that is capable of handling tensors of an arbitrary order.
    \item we developed a high performing implementation of the algorithm on Nvidia H100 GPUs leveraging their enormous performance, see~\Cref{sec:parallegpu,sec:implementation}.
    \item we evaluated the implementation on the arbitrary order and increasingly large tensors, yielding to good predictable scalability, see~\Cref{sec:results}, and compared against the Matlab reference implementation, showing more than two orders of magnitude speedup.
\end{itemize}

The remainder of the article is organized as follows: \Cref{sec:background} contains theoretical background plus an introduction to tensors, while \Cref{sec:jacobi} dives into our general Jacobi-type algorithm for tensor decomposition. \Cref{sec:implementation} provides an overview of tensor representation and our high performing implementation. \Cref{sec:results} presents convergence and performance results, and \Cref{sec:conclusion} discusses the main outcomes and outlines future work.

\section{Background and theory} 
\label{sec:background}

In this section, we present the theoretical background required to understand the algorithm under study and the developments that follow. 
%
We provide an overview of tensors and tensor operations that are fundamental to the algorithms and concepts discussed in this article. Unless stated otherwise, all the definitions concerning tensors found here align with the definitions presented in~\cite{kolda-review09}.

In this context, tensors can be thought of as multidimensional arrays that generalize vectors and matrices to higher orders. A vector is a 1-dimensional tensor and a matrix is a 2-dimensional tensor.
Tensors can be used to represent problems that depend on several significant variables. Therefore, they arise naturally in physics and scientific computing. Many-body quantum systems, high-dimensional simulations, and tensor‑structured operators produce data with multiple interacting dimensions. Tensors also appear in signal processing, where measurements may vary between sensors, time, and frequency, and in imaging tasks such as color images and video sequences. In these areas, tensor representations preserve the full multi‑dimensional structure of the problem without loss of information.

A tensor, or multidimensional array, is of order $D$ if it has $D$ dimensions (modes). Such a tensor can be denoted $\mathcal{X}\in\mathbb{R}^{I_1\times I_2 \times\cdots\times I_D}$, and, like vectors and matrices, it is indexed through $D$ indices as $x_{i_1i_2\ldots i_D}$. We work with \textit{cubical} tensors, which are those with $I_1=I_2=\cdots=I_D$, and we say that a cubical tensor is \textit{diagonal} if its only non-zero entries are those on positions $(i_1,i_2,\ldots,i_D)$, for $i_1=i_2=\cdots=i_D$.
\textit{Tensor matricization} refers to reorganizing a higher order tensor into a matrix by selecting mode $n$, $1\leq n\leq D$, as the row dimension and flattening all remaining modes into the columns according to some fixed, consistent ordering. This produces a mode‑$n$ unfolding that exposes the structure of the tensor along a single mode while preserving all entries. A matricized tensor can then be vectorized by flattening the unfolding into a single long vector.

Let $\mathcal{X}\in\mathbb{R}^{I_1\times\cdots\times I_n\times\cdots\times I_D}$ and $M\in\mathbb{R}^{J\times I_n}$. The \textit{mode-$n$ multiplication} is a tensor-matrix product
$$\mathcal{T}=\mathcal{X}\times_n M,$$
where $\mathcal{T}$ is a $I_1\times\cdots\times I_{n-1}\times J\times I_{n+1}\times\cdots\times I_D$ tensor, such that
\begin{equation}\label{eq:theory-tensor-mode-n-mult}
t_{i_1\ldots i_{n-1}ji_{n+1}\ldots i_D} = \sum_{i_n=1}^{I_n} x_{i_1\ldots i_n\ldots i_D}m_{ji_n}, \,\,  1\leq j\leq J.
\end{equation}
The \textit{Tucker decomposition} is a factorization of a tensor $\mathcal{X}\in\mathbb{R}^{I_1\times I_2\times\cdots\times I_D}$ into one core tensor $\mathcal{C}$ and $D$ factor matrices $M_1,M_2,\ldots,M_D$,
\begin{equation}
\mathcal{X}\approx\mathcal{C}\times_1M_1\times_2M_2\cdots\times_DM_D.
\end{equation}
A comprehensive overview of applications of the Tucker decomposition can be found in~\cite{kolda-review09}. 
A number of software libraries support Tucker decompositions of tensors of arbitrary dimensionality. Most of these libraries provide Tucker decomposition methods based on higher‑order singular value decomposition (HOSVD), which is a direct numerical method, or on higher‑order orthogonal iteration (HOOI), both originally introduced by De Lathauwer et al. \cite{Lathauwer00-2,Lathauwer00}. 

In addition, there are many libraries that offer building blocks for the implementation of Tucker decomposition methods such as HOSVD. For example, the following CUDA specific libraries are relevant in this context: cuBLAS, cuSOLVER, cuTensor, cuTensorNet. 

While Tucker decomposition libraries are widely available and provide efficient implementations of standard methods like HOSVD and HOOI, they are not designed to diagonalize the core tensor. In contrast, the Jacobi-type algorithm studied in this work explicitly seeks orthogonal transformations to make the core tensor as diagonal as possible. Because standard Tucker software leaves the core unconstrained and dense, it solves a fundamentally different optimization problem and cannot serve as a direct performance baseline for the diagonalization approach presented here

\section{Jacobi-type methods for tensor decomposition}
\label{sec:jacobi}

One of the main advantages of the Jacobi matrix algorithm is its inherent parallelism~\cite{LukPark89,DrmacVeselic07,Novakovic14}. Following the same idea, we develop a parallelization strategy for the Jacobi-type algorithm for diagonalization of the tensors of order-$D$. In this section, we describe the Jacobi-type diagonalization algorithm for tensors and present the overall approach to its acceleration on GPUs. 

\subsection{Third-order Jacobi diagonalization algorithm}

As a generalization of the well-studied Jacobi eigenvalue algorithm that diagonalizes symmetric/Hermitian matrices~\cite{GolubVanLoan_book,DrmacVeselic07}, we introduce the diagonalization algorithm for tensors. It was studied for symmetric tensors~\cite{ULC183,ULC19,ULC20} and for general tensors~\cite{begovic_kovac_convergence_2023,BegovicPerkovic24}. We work with general tensors and the algorithm developed in~\cite{begovic_kovac_convergence_2023} forms the basis of the work presented here. 

For $\mathcal{A}\in\mathbb{R}^{N\times N\times N}$, our goal is to obtain the Tucker decomposition
$$\mathcal{A}=\mathcal{C}\times_1U\times_2 V\times_3W,$$
where $U$, $V$, and $W$ are orthogonal matrices, and the core tensor $\mathcal{C}$ is diagonal, in the case of unitarily diagonalizable tensors, or as diagonal as possible for nondiagonalizable cases. Thus, our objective function takes the form
\begin{equation}\label{eq:f}
f(U,V,W)= \|\text{diag}(\mathcal{A}\times_1 U^{T} \times_2 V^{T} \times_3 W^{T})\|^{2}\rightarrow \max.
\end{equation}
The main idea of the algorithm is the following: For $\mathcal{A}\in\mathbb{R}^{N\times N\times N}$, in each iteration $k$, we apply to the underlying tensor three Givens rotations, one in each mode. The rotations are chosen to maximize the Frobenius norm of the tensor diagonal. We have
$$\mathcal{A}^{(k+1)}=\mathcal{A}^{(k)}\times_1R_{1,k}^T\times_2R_{2,k}^T\times_3R_{3,k}^T, \quad k\geq0.$$
Givens rotations $R_{i,k}$, $i=1,2,3$, differ from the identity matrix $I_N$ in a $2\times2$ submatrix
$$\begin{bmatrix}
\cos\phi_i^{(k)} & -\sin\phi_i^{(k)} \\
\sin\phi_i^{(k)} & \cos\phi_i^{(k)}
\end{bmatrix},$$
which is placed at the intersection of the $p(k)$th and $q(k)$th row and column of $R_{i,k}$. Therefore, in the $k$th iteration, rotations have the same pivot position $(p(k),q(k))$, but a different rotation angle. Pivot pairs are taken cyclically from the set $\{(p,q), 1\leq p<q\leq N\}$ .  

Rotation angles are chosen, one by one, by the alternating least squares approach. In other words, each iteration is made of three microiterations. In the first microiteration we consider only the change in the first mode, while the other two rotations are assumed to be identity, etc. Then, $\phi_1^{(k)}$ is the solution of the maximization problem
$$g(\phi)=(\mathcal{A}_{ppp}\cos\phi + \mathcal{A}_{qpp}\sin\phi)^{2} +
(-\mathcal{A}_{pqq}\sin\phi+\mathcal{A}_{qqq}\cos\phi)^{2}\rightarrow\max,$$
given by
\begin{equation}\label{eq:tan2phi}
\tan(2\phi) = \frac{2\lambda}{\mu},
\end{equation}
where
\begin{align}
\lambda & = 2(\mathcal{A}_{ppp}\mathcal{A}_{qpp}-\mathcal{A}_{pqq}\mathcal{A}_{qqq}) \mathrm{sign}(\mathcal{A}_{ppp}^{2}+\mathcal{A}_{qqq}^{2}-\mathcal{A}_{pqq}^{2}-\mathcal{A}_{qpp}^{2}),\label{eq:lambda} \\
\mu & = |\mathcal{A}_{ppp}^{2}+\mathcal{A}_{qqq}^{2}-\mathcal{A}_{pqq}^{2}-\mathcal{A}_{qpp}^{2}|. \label{eq:mu}
\end{align}
Note that it is not needed to compute $\phi$ explicitly, but to get $\sin\phi$ and $\cos\phi$ using~\eqref{eq:tan2phi}. Rotations in the second and the third mode are obtained analogously, see~\cite{begovic_kovac_convergence_2023} for details.

According to the convergence results from~\cite{begovic_kovac_convergence_2023}, a pivot pair is accepted if it satisfies the gradient-based condition\\
\begin{equation}\label{eq:pivot-check}
    2|(\Lambda(Q))_{pq}|\geq\eta\|\Lambda(Q)\|_2,
\end{equation}
for $Q=U,V,W$, in modes $1,2,3$, respectively.
Here, $\eta>0$ is a tuning parameter, $0<\eta\leq\frac{2}{N}$, and
\begin{equation}\label{eq:lambdaq}
\Lambda(Q) = \frac{Q^{T}\nabla_{Q}\tilde{f} - (\nabla_{Q}\tilde{f})^{T}Q}{2}
\end{equation}
is a projection of the function $\tilde{f}$, defined in the same way as $f$ from~\eqref{eq:f}, but for any square matrices $U,V,W$. Its gradient is calculated element-wise as 
\begin{equation}\label{eq:grad}
(\nabla_{U}\tilde{f})_{ml} = 2(\mathcal{A}\times_1U^T \times_2V^T \times_3W^T)_{lll}(\mathcal{A}\times_2V^T\times_3W^T)_{mll},
\end{equation}
in mode-$1$, and in the same way for modes $2$ and $3$.

\subsection{Generalization to order-$D$ tensors}
\label{sec:general}

This algorithm can be extended to tensors of arbitrary order $D\geq3$, however theoretical proofs are much more elaborate. For a tensor $\mathcal{T}\in\mathbb{R}^{N\times N\times\cdots\times N}$ of order $D$, the generalized method seeks a decomposition
$$\mathcal{T}=\mathcal{C}\times_1M_1\times_2M_2\cdots\times_DM_D,$$
such that $M_1,\ldots,M_D$ are orthogonal matrices and the order-$D$ core tensor $\mathcal{C}$ is as diagonal as possible. Now, one iteration consists of $D$ microiterations, which requires generalizing the quantities $\lambda$ and $\mu$ from the relations~\eqref{eq:lambda} and~\eqref{eq:mu}, respectively. We set the index‑replacement notation,
$$\mathcal{A}_{p\rightarrow q,n}\coloneqq\mathcal{A}_{p\ldots pqp\ldots p},$$
where the $n$th index is $q$, while the other indices are equal to $p$. Then,
\begin{align*}
\lambda & = 2(\mathcal{A}_{pp\ldots p}\mathcal{A}_{p\rightarrow q,n} - \mathcal{A}_{q\rightarrow p,n}\mathcal{A}_{qq\ldots q}) \mathrm{sign}(\mathcal{A}_{pp\ldots p}^{2}+\mathcal{A}_{qq\ldots q}^{2} - \mathcal{A}_{q\rightarrow p,n}^{2} - \mathcal{A}_{p\rightarrow q,n}^{2}), \\
\mu & = |\mathcal{A}_{pp\ldots p}^{2}+\mathcal{A}_{qq\ldots q}^{2} - \mathcal{A}_{q\rightarrow p,n}^{2} - \mathcal{A}_{p\rightarrow q,n}^{2}|.
\end{align*}
The gradient~\eqref{eq:grad} is generalized in the same manner,
\begin{equation}\label{eq:gradD}(\nabla_{M_1}\tilde{f})_{ml} = 2(\mathcal{A}\times_1M_1^T\cdots\times_DM_D^T)_{ll\cdots l} (\mathcal{A}\times_2M_2\cdots\times_DM_D)_{l\rightarrow m,1},
\end{equation}
in the first mode, etc. These expressions are a direct higher-order analogues of the third-order formulas, with the only difference being that the tensor now has $D$ indices rather than three. The higher-order structure is shown in~\Cref{alg:alg-pseudocode-2}.

\begin{algorithm}[htbp]
\caption{General higher-order Jacobi diagonalization algorithm.}
\label{alg:alg-pseudocode-2}
\DontPrintSemicolon
\SetKwProg{Procedure}{Procedure}{}{}
\SetKwFor{For}{for}{do}{end for}
\SetKwFor{While}{while}{do}{end while}
\SetKwIF{If}{ElseIf}{Else}{if}{then}{else if}{else}{end if}

\Procedure{diagonalize($\mathcal{A}\in\mathbb{R}^{N\times N\times\cdots\times N}$ of order $D$)}{
    $\mathcal{A}^{(0)}\gets\mathcal{B}\gets\mathcal{A}$\;
    $M_d^{(0)} \gets I \quad \text{for all } d = 1,\dots,D$\;
    $k=0$\;
    \While{not converged}{
        Choose pivot pair $(p,q)$\;

        \For{$m = 1$ \KwTo $D$}{
            \If{$(p,q)$ satisfies \eqref{eq:pivot-check}}{
                Find $\cos\phi_k$, $\sin\phi_k$ for $R_{m,k}$ using~\eqref{eq:tan2phi}\;

                $\mathcal{B}\gets\mathcal{B} \times_m R_{m,k}$\;
                $M_m^{(k+1)} \gets M_m^{(k)} R_{m,k}$\;
            }
        }

        $\mathcal{A}^{(k+1)} \gets \mathcal{B}$\;
    }
}
\end{algorithm}

\subsection{Parallelization strategy}
\label{sec:parallegpu}

The generalized Jacobi-type algorithm admits a natural form of parallelism. 
As shown in~\cite{begovic_kovac_convergence_2023}, convergence is guaranteed for any cyclic pivot strategy, 
meaning that pivot pairs may be processed in any fixed order across iterations. 
This flexibility allows the pivot pairs to be reorganized into groups that allow 
parallelism within each group.

For a given mode $m$, two pivot pairs $(p_1,q_1)$ and $(p_2,q_2)$ affect disjoint 
sets of tensor entries whenever
\[
p_1 \neq p_2,\qquad p_1 \neq q_2,\qquad q_1 \neq p_2,\qquad q_1 \neq q_2.
\]
If these conditions hold, the corresponding rotations commute and may be applied 
independently. This makes it possible to partition the full set of 
$\frac{N(N-1)}{2}$ pivot pairs into groups in which all pairs satisfy the 
disjointness condition. For even $N$, such a partition consists of $N-1$ groups; 
for odd $N$, it consists of $N$ groups. For example, when $N=4$, a valid 
partition is
\[
\{(1,2),(3,4)\},\qquad \{(1,3),(2,4)\},\qquad \{(1,4),(2,3)\}.
\]
\begin{algorithm}[htbp]
\caption{Parallel higher-order Jacobi diagonalization algorithm.}
\label{alg:alg-pseudocode-3}

\DontPrintSemicolon
\SetKwProg{Procedure}{Procedure}{}{}
\SetKwFor{For}{for}{do}{end for}
\SetKwFor{While}{while}{do}{end while}
\SetKwIF{If}{ElseIf}{Else}{if}{then}{else if}{else}{end if}

\Procedure{diagonalize($\mathcal{A}\in\mathbb{R}^{N\times N\times \cdots\times N}$ of order $D$)}{
    $\mathcal{A}^{(0)} \gets \mathcal{B} \gets \mathcal{A}$\;
    $M_d^{(0)} \gets I \quad \text{for all } d = 1,\dots,D$\;
    $k=0$\;
    \While{not converged}{
        Choose pivot group $G$\;

        \For{$m = 1$ \KwTo $D$}{
            $F \gets \{(p,q)\in G \mid (p,q)\text{ satisfies pivot check~\eqref{eq:pivot-check} for }Q=M_{m}\}$\;

            \ForEach{$(p,q)\in F$}{
                Compute $\cos\phi_k$ and $\sin\phi_k$ for $R_{m,k}$\;
            }

            Apply all rotations $\{R_{m,k} \mid (p,q)\in F\}$ to $\mathcal{B}$ and $M_m$\;
        }

        $\mathcal{A}^{(k+1)} \gets \mathcal{B}$\;
    }
}
\end{algorithm}

\section{High performing implementation}
\label{sec:implementation}
The CUDA implementation follows the overall structure outlined in~\Cref{alg:alg-pseudocode-3}, utilizing the inherent parallelism of GPUs. A key design choice is that no data, except the convergence check, is transferred between host and device during the execution of the algorithm.  Instead, all device buffers are allocated and the necessary data is copied from host to device prior to execution. The implementation then operates entirely in device memory, and results are moved back to the host once the algorithm has completed.

The implementation is exposed through a function named \texttt{diagonalize}, which accepts the input tensor and the initialized factor matrices.
Internally, this function allocates all required device memory buffers, and copies the input data to the device. The full list of device buffers is extensive and includes the input tensor, the factor matrices, the workspaces used by external library routines, the sines and cosines corresponding to the optimal rotation angles, and storage for intermediate results. Additionally, \texttt{diagonalize} also generates a pivot pair partitioning as explained in~\Cref{sec:general}, and copies that to the device memory.

After buffer allocation and data transfer, \texttt{diagonalize} implements the algorithm's main loop, which invokes three main functionalities

\begin{enumerate}
    \item \texttt{filter\_pq\_group} takes a group of pivot pairs and filters it so that only the pairs that pass the pivot check remain. Internally, it invokes a number of custom kernels, as well as a few cuBLAS, cuSOLVER, and CUB routines.

    \item \texttt{set\_rotations} takes a group of pivot pairs filtered by \texttt{filter\_pq\_group} and the core tensor at the current micro-iteration, and computes the optimal sine and cosine for each pivot pair.
    \item \texttt{apply\_rotations} takes the core tensor or the factor matrix, at the current micro-iteration, the filtered pivot pair group, and the corresponding sines and cosines as calculated by \texttt{set\_rotations}, and performs the update.
\end{enumerate}

The above three steps are performed for each iteration, for each pivot pair group, for each mode. Once the algorithm has completed, the core tensor and the factor matrices are copied back to the host.

\subsection{Tensor Representation}
In the implementation, a tensor is represented by a contiguous array of values (a vectorization) together with its order, a uniform dimension size, and an array of memory strides. This representation allows the tensor layout in memory to be fully described by the stride array, which provides flexibility in how multi‑indices map to linear indices.

The array of tensor values is stored in the global memory. The stride array is placed in the constant memory, since it is small and read‑only throughout the algorithm. The tensor order and dimension size are passed by value to kernels.

\subsection{Workload Partitioning}
All custom CUDA kernels use the same one-dimensional launch configuration. 
Blocks are arranged in a 1D grid, and threads within each block are also 1D. 
Given a problem size $S$ and a chosen number of threads per block $t_b$, the number 
of blocks is computed dynamically so that the total number of threads equals the 
largest power of two not exceeding $S$, divided by $t_b$. This ensures that no 
thread is assigned more than two elements of the workload.

Each thread identifies itself by a global thread index
($i = b_{\text{id}} \cdot t_b + t_{\text{id}}$)
and uses this index to determine the bounds of its assigned sub-problem. 
The total workload of size $S$ is partitioned into 
$p = b_g \cdot t_b$ contiguous chunks that differ in size by at most one element. 
Each thread computes the first and last index of its chunk using a standard linear 
partitioning scheme, ensuring full coverage of the workload without overlap and 
a per-thread workload that is as balanced as possible.

This strategy is used consistently across the kernels in the implementation.

\subsection{Core Functions}
\paragraph{Filtering Pivot Pairs.} 
Filtering of the current pivot pair group is arguably the most complex part of the entire implementation.
Both in terms of its representation in code and in terms of computational cost.
Internally, \texttt{filter\_pq\_group} consists of multiple calls to custom built kernels, and to external CUDA libraries like cuBLAS, cuSOLVER, and CUB (all of which are available through the CUDA Toolkit \cite{nvidia_cuda_toolkit}). For a given mode $m$, the current core tensor $A^{(k)}$, and the current group of pivot pairs, the following steps outline its implementation:

\begin{enumerate}
    \item {\em Calculation of $\Lambda( Q)$}. This is the most complex part of \texttt{filter\_pq\_group}. Given the current core tensor $\mathcal{A}^{(k)}$ and the current factor matrices $U_1^{(k)},U_2^{(k)},$ $\ldots, U_D^{(k)}$, it involves calculation of $\nabla_Q \tilde{f}$ as defined by~\eqref{eq:gradD}, which in turn requires reconstruction of the two tensors,
    \[
        \begin{aligned}
                \mathcal{T}_1 & = \mathcal{A}^{(k)} \times _1 U_1^{(k)} \times _2 U_2^{(k)} \cdots \times _D U_D^{(k)}, \\
                \mathcal{T}_2 & = \mathcal{A}^{(k)} \times _1 U_1^{(k)} \times _{m-1} U_{m-1}^{(k)} \times_{m+1} U_{m+1}^{(k)}\cdots \times _D U_D^{(k)}.
        \end{aligned}
    \]
Construction of the two full tensors is implemented as a sequential loop over the modes, running on the CPU, with each iteration invoking a custom built CUDA kernel implementing element-wise mode-n multiplication.
Using the two tensors, another custom built kernel then produces $\nabla_Q \tilde{f}$. Together with the factor matrix $Q$, $\nabla_Q \tilde{f}$ then forms $\Lambda(Q)$ through a series of calls to cuBLAS routines for matrix-matrix multiplication according to  \eqref{eq:lambdaq}.

\item {\em Calculation of $||\Lambda(Q)||_2$}.
After construction of $\Lambda(Q)$, its 2-norm $||\Lambda(Q)||_2$ is needed to perform the actual pivot checks. Since the 2-norm, at least in this context, is equivalent to the largest singular value of the matrix, this step is implemented using SVD routines available through cuSOLVER.
    
\item {\em Filtering based on $\Lambda(Q)$ and $||\Lambda(Q)||_2$}.
After calculation of $\Lambda(Q)$ and its 2-norm, the current group of pivot pairs is then filtered according to the condition in \eqref{eq:pivot-check}, using a mark-compact approach. This involves computing an array of flags where a flag is set if and only if the corresponding pivot pair passed the pivot check. The flag array is then used to calculate an array of prefix sums that tell us how the group should be compacted to remove the pairs that did not pass the check. Calculation of the prefix sum array is implemented using CUB routines. Finally, the prefix sum array is used to compact the pivot pair group, using a custom kernel.
\end{enumerate}

After passing a pivot pair group through the above steps, it contains only the pivot pairs that pass the pivot check, and it is then ready to be used by {\tt set\_rotation} and {\tt apply\_rotations}.

\paragraph{Application of Rotations.} For a given mode and pivot--pair group, the application of rotations is parallelized over tensor entries. According to the workload-partitioning strategy, each thread executing \texttt{apply\_rotations} is assigned at most two tensor multi--indices, which provides a fixed and predictable amount of work per thread regardless of tensor order or dimension size.

Each thread processes its assigned multi-indices independently. For each multi-index, it extracts the $n$th coordinate $i_n$. The filtered pivot-pair group for mode $n$ is stored as a list whose length does not exceed the dimension size $N$, which allows $i_n$ to be used directly as an index into this list.

If $i_n$ is smaller than the size of the filtered pivot-pair group, the corresponding rotation parameters are retrieved and the associated Givens rotation is applied to that specific tensor entry. If $i_n$ falls outside the range of the pivot-pair list, no update is performed for that entry. Apart from this bounds check, the kernel performs no additional branching. \Cref{lst:apply-rotations} outlines the procedure.

\begin{lstlisting}[language=C,caption={Sketch of the \texttt{apply\_rotations} kernel to update the core tensor.},label={lst:apply-rotations}]
__global__ void apply_rotations(float* tensor, int tensor_size, int N, 
     const PQPair* group, int group_size, const float* s_values, 
     const float c_values, int mode) 
{
    int start, end;
    if (!compute_thread_bounds(tensor_size, &start, &end))
        return;

    for (int i = start; i < end; ++i) {
        int in = get_mode_n_index(i, mode, N);          // Extract n-th coordinate

        if (in >= group_size)                  // Check whether a rotation applies
            continue;

        float c = c_values[in];                // Retrieve rotation and pivot pair
        float s = -s_values[in];
        PQPair pq = group[in];

        int base = set_mode_n_index(i, in, 0, mode);     // Compute linear indices
        int idx_p = set_mode_n_index(base, 0, pq.p, mode);  
        int idx_q = set_mode_n_index(base, 0, pq.q, mode);

        float Tp = tensor[idx_p];
        float Tq = tensor[idx_q];
        tensor[idx_q] = s * Tp + c * Tq;                  // Apply Givens rotation
        tensor[idx_p] = c * Tp - s * Tq;
} }
\end{lstlisting}

\section{Experimental results}
\label{sec:results}

We conducted measurements of the MATLAB version, the sequential C-version, and the CUDA versions on the Pelle cluster at UPPMAX using Nvidia H100 GPUs. We examined the convergence and runtime behavior of the final CUDA implementation using diagonalizable tensors with progressively increasing order and dimension sizes. We also measured the end‑to‑end speedup achieved by the CUDA version in comparison with the original MATLAB implementation.

\subsection{Convergence behavior}
\label{sec:convergence-behavior}
Here we present the convergence behavior of the final CUDA version, when applied to diagonalizable tensors of different sizes. We look at the convergence of the relative off-diagonal norm of the core tensor
$$\text{off}^2(\mathcal{X})\coloneqq\|\mathcal{X}\|_F^2-\|\text{diag}(\mathcal{X})\|_F^2$$
under two types of scaling, separately; scaling of the number of dimensions, and scaling of the dimension size. All diagonalizable tensors were generated according to the following steps:
\begin{itemize}
    \item[(i)] Generate a diagonal tensor $\mathcal{C}$ with random values on the diagonal.
    \item[(ii)] Generate $D$ random unitary matrices $M_1,M_2,\ldots,M_D$.
    \item[(iii)] Form $\mathcal{T}$ as $\mathcal{T}=\mathcal{C}\times_1M_1\times_2M_2\cdots\times_DM_D$.
    \item[(iv)] {\em Verification:} We apply the implementation to $\mathcal{T}$ and verify its correctness by comparing the obtained results to the original components.
\end{itemize}

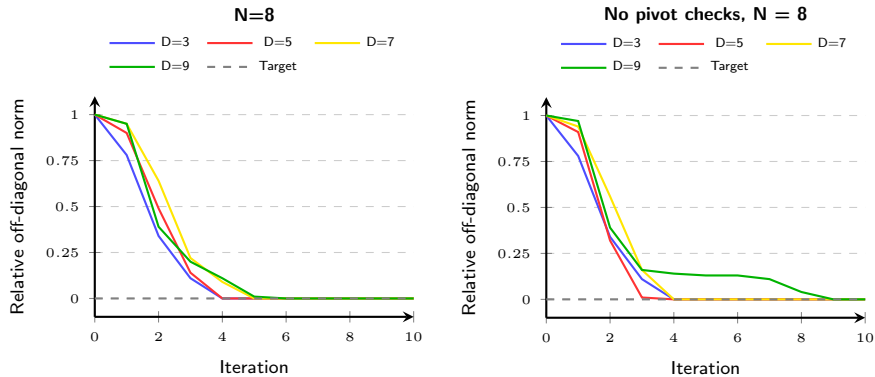
\begin{figure}
\begin{minipage}{0.48\textwidth}
\centering
\begin{tikzpicture}
\begin{axis}[
    title={\textbf{N=8}},
    title style={align=center, at={(0.5,1.3)}, anchor=center, font=\sffamily\scriptsize},
    xlabel={Iteration},
    ylabel={Relative off-diagonal norm},
    xmin=0, xmax=10,
    ymin=-0.1, ymax=1.1,
    xtick={0,2,4,6,8,10}, 
    ytick={0, 0.25, 0.50, 0.75, 1.00},
    ymajorgrids=true,             
    grid style={dashed, gray!40},
    tick label style={/pgf/number format/fixed, /pgf/number format/precision=2, font=\sffamily\tiny},
    axis x line=bottom,
    axis y line=left,
    enlarge x limits=false,
    legend style={
        at={(0.5,1.05)},
        anchor=south,
        legend columns=3, 
        draw=none,
        font=\sffamily\tiny,
        /tikz/every even column/.append style={column sep=0.15cm}
    },
    label style={font=\sffamily\scriptsize},
    width=5.8cm,  
    height=4.5cm, 
    thick
]

\addplot[color=blue!70!white, mark=none, thick] coordinates {
    (0, 1.00) (1, 0.78) (2, 0.34) (3, 0.11) (4, 0.00) (5, 0.00) (6, 0.00) (7, 0.00) (8, 0.00) (9, 0.00) (10, 0.00)
};
\addlegendentry{D=3}

\addplot[color=red!80!white, mark=none, thick] coordinates {
    (0, 1.00) (1, 0.90) (2, 0.49) (3, 0.14) (4, 0.00)
    (5, 0.00) (6, 0.00) (7, 0.00) (8, 0.00) (9, 0.00) (10, 0.00)
};
\addlegendentry{D=5}

\addplot[color=yellow!90!orange, mark=none, thick] coordinates {
    (0, 1.00) (1, 0.95) (2, 0.64) (3, 0.22) (4, 0.09)
    (5, 0.00) (6, 0.00) (7, 0.00) (8, 0.00) (9, 0.00) (10, 0.00)
};
\addlegendentry{D=7}

\addplot[color=green!70!black, mark=none, thick] coordinates {
    (0, 1.00) (1, 0.95) (2, 0.39) (3, 0.20) (4, 0.11)
    (5, 0.01) (6, 0.00) (7, 0.00) (8, 0.00) (9, 0.00) (10, 0.00)
};
\addlegendentry{D=9}

\addplot[color=gray, dashed, thick] coordinates {
    (0, 0) (10, 0)
};
\addlegendentry{Target}

\end{axis}
\end{tikzpicture}
\end{minipage}
\begin{minipage}{0.48\textwidth}
\centering
\begin{tikzpicture}
\begin{axis}[
    title={\textbf{No pivot checks, N = 8}},
    title style={align=center, at={(0.5,1.3)}, anchor=center, font=\sffamily\scriptsize},
    xlabel={Iteration},
    ylabel={Relative off-diagonal norm},
    xmin=0, xmax=10,
    ymin=-0.1, ymax=1.1,
    xtick={0,2,4,6,8,10}, 
    ytick={0, 0.25, 0.50, 0.75, 1.00},
    ymajorgrids=true,             
    grid style={dashed, gray!40},
    tick label style={/pgf/number format/fixed, /pgf/number format/precision=2, font=\sffamily\tiny},
    axis x line=bottom,
    axis y line=left,
    enlarge x limits=false,
    legend style={
        at={(0.5,1.05)},
        anchor=south,
        legend columns=3, 
        draw=none,
        font=\sffamily\tiny,
        /tikz/every even column/.append style={column sep=0.15cm}
    },
    label style={font=\sffamily\scriptsize},
    width=5.8cm,  
    height=4.5cm, 
    thick
]

\addplot[color=blue!70!white, mark=none, thick] coordinates {
    (0, 1.00) (1, 0.78) (2, 0.34) (3, 0.11) (4, 0.00) (5, 0.00) (6, 0.00) (7, 0.00) (8, 0.00) (9, 0.00) (10, 0.00)
};
\addlegendentry{D=3}

\addplot[color=red!80!white, mark=none, thick] coordinates {
    (0, 1.00) (1, 0.91) (2, 0.32) (3, 0.01) (4, 0.00)
    (5, 0.00) (6, 0.00) (7, 0.00) (8, 0.00) (9, 0.00) (10, 0.00)
};
\addlegendentry{D=5}

\addplot[color=yellow!90!orange, mark=none, thick] coordinates {
    (0, 1.00) (1, 0.94) (2, 0.56) (3, 0.16) (4, 0.00)
    (5, 0.00) (6, 0.00) (7, 0.00) (8, 0.00) (9, 0.00) (10, 0.00)
};
\addlegendentry{D=7}

\addplot[color=green!70!black, mark=none, thick] coordinates {
    (0, 1.00) (1, 0.97) (2, 0.39) (3, 0.16) (4, 0.14)
    (5, 0.13) (6, 0.13) (7, 0.11) (8, 0.04) (9, 0.00) (10, 0.00)
};
\addlegendentry{D=9}

\addplot[color=gray, dashed, thick] coordinates {
    (0, 0) (10, 0)
};
\addlegendentry{Target}
\end{axis}
\end{tikzpicture}
\end{minipage}
    \caption{Convergence of the relative off-diagonal norm on four different diagonalizable tensors with dimension size  $N=8$ and orders $D=3, 5, 7, 9$: (left) with pivot checks; (right) no checks. All measurements, except for the one corresponding to the dashed blue line, were generated using $\eta=0.00125$.}
    \label{fig:convergence-diagonalizable-n8}
\end{figure}

\Cref{fig:convergence-diagonalizable-n8} (left)
shows convergence behavior with pivot checks enabled. The pivot check mathematically guarantees convergence 
for $0<\eta\leq\frac{2}{N}$.  Increasing $\eta$  accelerates convergence, but it also causes the algorithm to settle at a solution that is less accurate than the one obtained for smaller values. An example of the effect of the choice of $\eta$ is shown in~\Cref{fig:convergence-diagonalizable-n8} (left) where, given the specific problem of size 
$N=8$ and $D=3$, the relative off-diagonal norm converges to its initial value for $\eta = 0.00125$, and to approximately zero (the expected solution) for $\eta=0.000125$. Furthermore, by this logic, setting $\eta$ to equal zero (which is equivalent to completely disabling pivot checks) should yield the most exact solution possible. The problem is just that, for this case, convergence is not mathematically proven. 

\begin{figure}
\begin{minipage}{0.48\textwidth}
\centering
\begin{tikzpicture}
\begin{axis}[
    title={\textbf{D=3}},
    title style={align=center, at={(0.5,1.3)}, anchor=center, font=\sffamily\scriptsize},
    xlabel={Iteration},
    ylabel={Relative off-diagonal norm},
    xmin=0, xmax=10,
    ymin=-0.1, ymax=1.1,
    xtick={0,2,4,6,8,10},
    ytick={0, 0.25, 0.50, 0.75, 1.00},
    ymajorgrids=true,             
    grid style={dashed, gray!40},
    tick label style={/pgf/number format/fixed, /pgf/number format/precision=2, font=\sffamily\tiny},
    axis x line=bottom,
    axis y line=left,
    enlarge x limits=false,
    legend style={
        at={(0.5,1.05)},
        anchor=south,
        legend columns=3, 
        draw=none,
        font=\sffamily\tiny,
        /tikz/every even column/.append style={column sep=0.15cm}
    },
    label style={font=\sffamily\scriptsize},
    width=5.8cm,  
    height=4.5cm, 
    thick
]

\addplot[color=blue!70!white, mark=none, thick] coordinates {
    (0, 1.00) (1, 0.81) (2, 0.36) (3, 0.09) (4, 0.01)
    (5, 0.00) (6, 0.00) (7, 0.00) (8, 0.00) (9, 0.00) (10, 0.00)
};
\addlegendentry{N=32}

\addplot[color=red!80!white, mark=none, thick] coordinates {
    (0, 1.00) (1, 0.91) (2, 0.52) (3, 0.18) (4, 0.07)
    (5, 0.02) (6, 0.00) (7, 0.00) (8, 0.00) (9, 0.00) (10, 0.00)
};
\addlegendentry{N=64}

\addplot[color=yellow!90!orange, mark=none, thick] coordinates {
    (0, 1.00) (1, 0.90) (2, 0.46) (3, 0.19) (4, 0.07)
    (5, 0.03) (6, 0.00) (7, 0.00) (8, 0.00) (9, 0.00) (10, 0.00)
};
\addlegendentry{N=128}

\addplot[color=green!70!black, mark=none, thick] coordinates {
    (0, 1.00) (1, 0.90) (2, 0.48) (3, 0.19) (4, 0.07)
    (5, 0.03) (6, 0.02) (7, 0.01) (8, 0.00) (9, 0.00) (10, 0.00)
};
\addlegendentry{N=256}

\addplot[color=gray, dashed, thick] coordinates {
    (0, 0) (10, 0)
};
\addlegendentry{Target}
\end{axis}
\end{tikzpicture}
\end{minipage}
\begin{minipage}{0.48\textwidth}
\centering
\begin{tikzpicture}
\begin{axis}[
    title={\textbf{No pivot checks, D = 3}},
    title style={align=center, at={(0.5,1.3)}, anchor=center, font=\sffamily\scriptsize},
    xlabel={Iteration},
    ylabel={Relative off-diagonal norm},
    xmin=0, xmax=10,
    ymin=-0.1, ymax=1.1,
    xtick={0,2,4,6,8,10},
    ytick={0, 0.25, 0.50, 0.75, 1.00},
    ymajorgrids=true,             
    grid style={dashed, gray!40},
    tick label style={/pgf/number format/fixed, /pgf/number format/precision=2, font=\sffamily\tiny},
    axis x line=bottom,
    axis y line=left,
    enlarge x limits=false,
    legend style={
        at={(0.5,1.05)},
        anchor=south,
        legend columns=3,
        draw=none,
        font=\sffamily\tiny,
        /tikz/every even column/.append style={column sep=0.15cm}
    },
    label style={font=\sffamily\scriptsize},
    width=5.8cm,  
    height=4.5cm, 
    thick
]

\addplot[color=blue!70!white, mark=none, thick] coordinates {
    (0, 1.00) (1, 0.81) (2, 0.36) (3, 0.09) (4, 0.01)
    (5, 0.00) (6, 0.00) (7, 0.00) (8, 0.00) (9, 0.00) (10, 0.00)
};
\addlegendentry{N=32}

\addplot[color=red!80!white, mark=none, thick] coordinates {
    (0, 1.00) (1, 0.91) (2, 0.52) (3, 0.18) (4, 0.07)
    (5, 0.02) (6, 0.00) (7, 0.00) (8, 0.00) (9, 0.00) (10, 0.00)
};
\addlegendentry{N=64}

\addplot[color=yellow!90!orange, mark=none, thick] coordinates {
    (0, 1.00) (1, 0.90) (2, 0.46) (3, 0.19) (4, 0.07)
    (5, 0.03) (6, 0.00) (7, 0.00) (8, 0.00) (9, 0.00) (10, 0.00)
};
\addlegendentry{N=128}

\addplot[color=green!70!black, mark=none, thick] coordinates {
    (0, 1.00) (1, 0.90) (2, 0.48) (3, 0.19) (4, 0.07)
    (5, 0.03) (6, 0.02) (7, 0.01) (8, 0.00) (9, 0.00) (10, 0.00)
};
\addlegendentry{N=256}

\addplot[color=gray, dashed, thick] coordinates {
    (0, 0) (10, 0)
};
\addlegendentry{Target}

\end{axis}
\end{tikzpicture}
\end{minipage}
    \caption{Convergence of the relative off-diagonal norm on four different diagonalizable tensors with order $D=3$ and dimension sizes $N=32, 64, 128, 256$: (left) with pivot checks; (right) no checks. All measurements were generated using $\eta = 0.00125$.}
    \label{fig:convergence-diagonalizable-d3}
\end{figure}
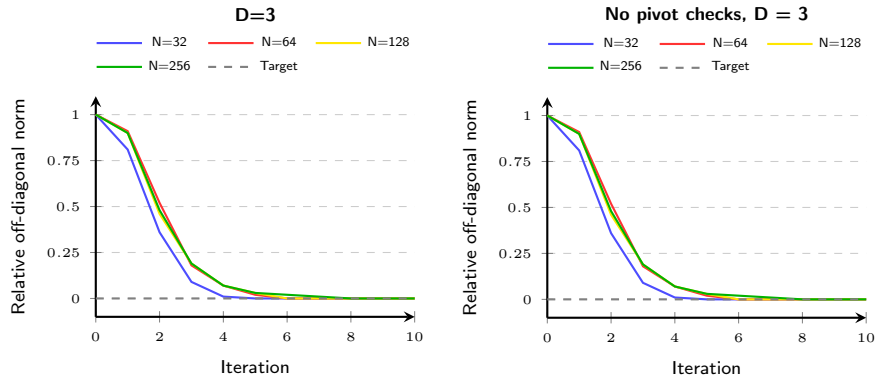

Interestingly, the evaluation presented here, performed on diagonalizable tensors, indicates that the algorithm converges even when $\eta$ is set to zero, \Cref{fig:convergence-diagonalizable-n8} (right).  \Cref{fig:convergence-diagonalizable-n8} shows the convergence behavior when increasing dimensionality, while \Cref{fig:convergence-diagonalizable-d3} for the behavior when increasing dimension size. Both graphs show how the relative off-diagonal norm converges to zero even with pivot checks disabled.
Additionally, supporting the theory that $\eta = 0$ should be optimal for performance (if convergence is achieved), when analyzing the accuracy of the resulting decompositions, as described in the verification step above, all results matched expectations up to the numerical tolerance ($\epsilon=10^{-7}$). These results together could have highly positive implications for the usability of the implementation, because as shown in~\Cref{sec:performance-experiments}, the pivot checks are computationally expensive.


\subsection{Performance evaluation}
\label{sec:performance-experiments}
Here we present performance results of the CUDA version (including the optimization steps) and the sequential C version, which served as  a transitional point from Matlab to CUDA. We used the same diagonalizable tensors for which convergence is demonstrated in~\Cref{sec:convergence-behavior}. Since all convergence results indicate that the implementation converges both with and without pivot checks, we evaluated runtime performance for both cases. Additionally, all measurements presented here represent 10 full cycles of the algorithm. A fixed number of iterations was used to ensure that we actually show scaling behavior rather than behavior induced by the specific problem, and 10 was chosen to ensure convergence; as demonstrated in~\Cref{sec:convergence-behavior}, convergence was achieved within 10 iterations for all evaluated tensors. 

For the CUDA versions we present elapsed GPU time, meaning that the corresponding measurements do not include time for memory allocation on host or device, data movement, kernel launch overhead, or execution on the CPU. Instead, they isolate the cost of the kernels, which dominates the total runtime. In a similar manner, the measurements corresponding to the CPU-version exclude memory allocation and setup, but here we measured wall-time over the implementation's main loop, when running with a single CPU thread. Note that the sequential C-version is utilized here to demonstrate relative acceleration and scaling behavior rather than as an exhaustive multi-core CPU baseline. All measurements were made on the Pelle cluster with two 48-core AMD EPYC 9454P (Zen4) @2.75\,GHz and two H100 GPUs per node, but one was utilized.

\begin{figure}
\begin{minipage}{0.48\textwidth}
\centering
\begin{tikzpicture}
\begin{axis}[
    title={\textbf{N=8}},
    title style={align=center, at={(0.5,1.35)}, anchor=center, font=\sffamily\scriptsize},
    xlabel={Order (D)},
    ylabel={Time in seconds},
    ymode=log,
    xmin=2.5, xmax=9.5, 
    ymin=0.01, ymax=1000, 
    xtick={3,4,5,6,7,8,9},
    ytick={0.01, 0.1, 1, 10, 100, 1000},
    yticklabels={$10^{-2}$, $10^{-1}$, $10^0$, $10^1$, $10^2$, $10^3$},
    ymajorgrids=true,             
    grid style={dashed, gray!40}, 
    axis x line=bottom,
    axis y line=left,
    enlarge x limits=false,
    legend style={
        at={(0.5,1.05)},
        anchor=south,
        legend columns=4,
        draw=none,
        font=\sffamily\tiny,
        /tikz/every even column/.append style={column sep=0.15cm}
    },
    label style={font=\sffamily\scriptsize},
    tick label style={font=\sffamily\tiny},
    width=5.8cm,  
    height=4.5cm, 
    thick
]

\addplot[color=blue!70!white, mark=none, thick] coordinates {
    (3, 0.083691) (4, 0.103209) (5, 0.121529) (6, 0.226154) (7, 1.349148) (8, 13.954719) (9, 133.569078)
};
\addlegendentry{CUDA 1}

\addplot[color=red!80!white, mark=none, thick] coordinates {
    (3, 0.079562) (4, 0.093905) (5, 0.108174) (6, 0.168592) (7, 0.740638) (8, 6.935002) (9, 70.980930)
};
\addlegendentry{CUDA 2}

\addplot[color=yellow!90!orange, mark=none, thick] coordinates {
    (3, 0.072255) (4, 0.083364) (5, 0.094782) (6, 0.125077) (7, 0.350241) (8, 2.786550) (9, 27.809240)
};
\addlegendentry{CUDA 3}

\addplot[color=green!70!black, mark=none, thick] coordinates {
    (3, 0.034946) (4, 0.357174) (5, 4.944761) (6, 59.704557) (7, 708.228994)
};
\addlegendentry{C}

\end{axis}
\end{tikzpicture}
\end{minipage}
\begin{minipage}{0.48\textwidth}
\centering
\begin{tikzpicture}
\begin{axis}[
    title={\textbf{No pivot checks, N=8}},
    title style={align=center, at={(0.5,1.35)}, anchor=center, font=\sffamily\scriptsize},
    xlabel={Order (D)},
    ylabel={Time in seconds},
    ymode=log,
    xmin=2.5, xmax=9.5, 
    ymin=0.0001, ymax=10000, 
    xtick={3,4,5,6,7,8,9},
    ytick={0.0001, 0.001, 0.01, 0.1, 1, 10, 100, 1000, 10000},
    yticklabels={$10^{-4}$, $10^{-3}$, $10^{-2}$, $10^{-1}$, $10^0$, $10^1$, $10^2$, $10^3$, $10^4$},
    ymajorgrids=true,             
    grid style={dashed, gray!40}, 
    axis x line=bottom,
    axis y line=left,
    enlarge x limits=false,
    legend style={
        at={(0.5,1.05)},
        anchor=south,
        legend columns=4,
        draw=none,
        font=\sffamily\tiny,
        /tikz/every even column/.append style={column sep=0.15cm}
    },
    label style={font=\sffamily\scriptsize},
    tick label style={font=\sffamily\tiny},
    width=5.8cm,  
    height=4.5cm, 
    thick
]

\addplot[color=blue!70!white, mark=none, thick] coordinates {
    (3, 0.007075) (4, 0.009510) (5, 0.012181) (6, 0.015398) (7, 0.034607) (8, 0.233554) (9, 1.947736)
};
\addlegendentry{CUDA 1}

\addplot[color=red!80!white, mark=none, thick] coordinates {
    (3, 0.004173) (4, 0.005439) (5, 0.006697) (6, 0.009073) (7, 0.022114) (8, 0.136742) (9, 1.163695)
};
\addlegendentry{CUDA 2}

\addplot[color=yellow!90!orange, mark=none, thick] coordinates {
    (3, 0.002874) (4, 0.003902) (5, 0.004995) (6, 0.006755) (7, 0.016143) (8, 0.097370) (9, 0.797681)
};
\addlegendentry{CUDA 3}

\addplot[color=green!70!black, mark=none, thick] coordinates {
    (3, 0.000713) (4, 0.007956) (5, 0.092179) (6, 1.083580) (7, 10.814357) (8, 114.194255) (9, 1133.994782)
};
\addlegendentry{C}

\end{axis}
\end{tikzpicture}
\end{minipage}
    \caption{Runtime in seconds of the different implementation versions when keeping dimension size constant ($N=8$) while varying the tensor order ($D=3,4,5,6,7,8,9$): (left) with pivot checks; (right) no checks.}
    \label{fig:runtimes-diagonalizable-n8}
\end{figure}
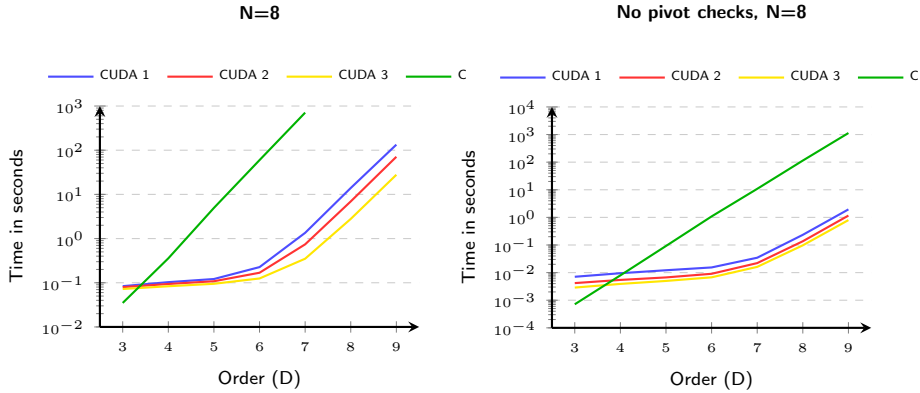

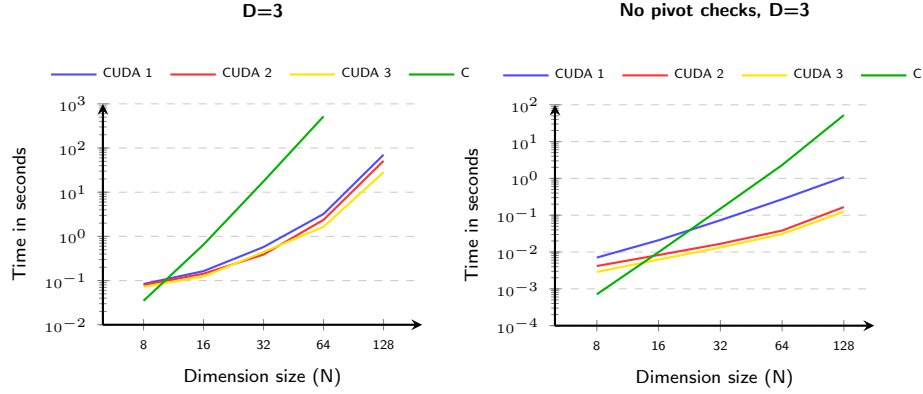
\begin{figure}
\begin{minipage}{0.48\textwidth}
\centering
\begin{tikzpicture}
\begin{axis}[
    title={\textbf{D=3}},
    title style={align=center, at={(0.5,1.35)}, anchor=center, font=\sffamily\scriptsize},
    xlabel={Dimension size (N)},
    ylabel={Time in seconds},
    xmode=log, 
    ymode=log,
    xmin=5, xmax=200, 
    ymin=0.01, ymax=1000, 
    xtick={8, 16, 32, 64, 128},
    xticklabels={8, 16, 32, 64, 128},
    ytick={0.01, 0.1, 1, 10, 100, 1000},
    yticklabels={$10^{-2}$, $10^{-1}$, $10^0$, $10^1$, $10^2$, $10^3$},
    ymajorgrids=true,             
    grid style={dashed, gray!40}, 
    axis x line=bottom,
    axis y line=left,
    enlarge x limits=false,
    legend style={
        at={(0.5,1.05)},
        anchor=south,
        legend columns=4,
        draw=none,
        font=\sffamily\tiny,
        /tikz/every even column/.append style={column sep=0.15cm}
    },
    label style={font=\sffamily\scriptsize},
    tick label style={font=\sffamily\tiny},
    width=5.8cm,  
    height=4.5cm, 
    thick
]

\addplot[color=blue!70!white, mark=none, thick] coordinates {
    (8, 0.083691) (16, 0.163847) (32, 0.575303) (64, 3.221361) (128, 70.514172)
};
\addlegendentry{CUDA 1}

\addplot[color=red!80!white, mark=none, thick] coordinates {
    (8, 0.079562) (16, 0.143194) (32, 0.38895) (64, 2.372148) (128, 51.195)
};
\addlegendentry{CUDA 2}

\addplot[color=yellow!90!orange, mark=none, thick] coordinates {
    (8, 0.072255) (16, 0.12368) (32, 0.443783) (64, 1.659189) (128, 28.273055)
};
\addlegendentry{CUDA 3}

\addplot[color=green!70!black, mark=none, thick] coordinates {
    (8, 0.034946) (16, 0.653419) (32, 17.629337) (64, 520.33422)
};
\addlegendentry{C}

\end{axis}
\end{tikzpicture}
\end{minipage}
\begin{minipage}{0.48\textwidth}
\centering
\begin{tikzpicture}
\begin{axis}[
    title={\textbf{No pivot checks, D=3}},
    title style={align=center, at={(0.5,1.35)}, anchor=center, font=\sffamily\scriptsize},
    xlabel={Dimension size (N)},
    ylabel={Time in seconds},
    xmode=log, 
    ymode=log,
    xmin=5, xmax=180, 
    ymin=0.0001, ymax=100, 
    xtick={8, 16, 32, 64, 128},
    xticklabels={8, 16, 32, 64, 128},
    ytick={0.0001, 0.001, 0.01, 0.1, 1, 10, 100},
    yticklabels={$10^{-4}$, $10^{-3}$, $10^{-2}$, $10^{-1}$, $10^0$, $10^1$, $10^2$},
    ymajorgrids=true,             
    grid style={dashed, gray!40}, 
    axis x line=bottom,
    axis y line=left,
    enlarge x limits=false,
    legend style={
        at={(0.5,1.05)},
        anchor=south,
        legend columns=4,
        draw=none,
        font=\sffamily\tiny,
        /tikz/every even column/.append style={column sep=0.15cm}
    },
    label style={font=\sffamily\scriptsize},
    tick label style={font=\sffamily\tiny},
    width=5.8cm,  
    height=4.5cm, 
    thick
]

\addplot[color=blue!70!white, mark=none, thick] coordinates {
    (8, 0.007075) (16, 0.020951) (32, 0.073131) (64, 0.271053) (128, 1.083583)
};
\addlegendentry{CUDA 1}

\addplot[color=red!80!white, mark=none, thick] coordinates {
    (8, 0.004173) (16, 0.008289) (32, 0.016741) (64, 0.038305) (128, 0.166302)
};
\addlegendentry{CUDA 2}

\addplot[color=yellow!90!orange, mark=none, thick] coordinates {
    (8, 0.002874) (16, 0.006238) (32, 0.013274) (64, 0.030902) (128, 0.124943)
};
\addlegendentry{CUDA 3}

\addplot[color=green!70!black, mark=none, thick] coordinates {
    (8, 0.000713) (16, 0.009936) (32, 0.150162) (64, 2.303391) (128, 52.297165)
};
\addlegendentry{C}

\end{axis}
\end{tikzpicture}
\end{minipage}
    \caption{Runtime in seconds of the different implementation versions (without pivot checks) when keeping tensor order constant ($D=3$) while varying the dimension size ($N=8, 16, 32, 64, 128$): (left) with pivot checks; (right) no checks.}
    \label{fig:runtimes-diagonalizable-d3}
\end{figure}

\paragraph{Scaling}.
\Cref{fig:runtimes-diagonalizable-n8,fig:runtimes-diagonalizable-d3} (left) present scaling with pivot checks, while the same figures (right) show scaling without pivot checks. In these figures, `CUDA 1' corresponds to the initial CUDA version; `CUDA 2' to the version that follows a new representation of tensors, which makes it possible to remove recurring calculation of tensor strides; `CUDA 3' to the final CUDA version that leverages the CUDA GPU's constant memory to achieve faster and more effective memory access.

The runtime of the sequential C-version increases (approximately) exponentially both when the tensor order increases by one (see  \Cref{fig:runtimes-diagonalizable-n8}), and when the order is kept constant and the dimension size increases by a factor of two (see \Cref{fig:runtimes-diagonalizable-d3}). However, the CUDA versions show much better scaling up to a certain point, at which they seem to gradually transition to scale in ways similar to the CPU version.
This transition is expected, since the GPU has limited parallel resources too. When the parallel resources become saturated, additional work can no longer be parallelized. Beyond this saturation point, computations inevitably become increasingly sequential.
To conclude the overall scaling behavior, we can say that the runtime of the CUDA implementation (approximately) scales exponentially with tensor order, and linearly in tensor size, as problems get large.

Another important observation is that when running with pivot checks, the pivot check itself dominates the runtime. For example, for the tensor of size $8^9$, \Cref{fig:runtimes-diagonalizable-n8} (right) shows a measurement of around 0.8 seconds (for the final CUDA version), while the corresponding measurement in~\Cref{fig:runtimes-diagonalizable-n8} (left) is approximately 28 seconds, which 35 times higher.

\paragraph{Speedup}. We also present the end-to-end speedup achieved by the final CUDA version without pivot checks compared to the MATLAB implementation. This comparison is limited to 3D tensors, since the MATLAB code does not support higher orders. 
Here, the end-to-end runtime is defined as the time elapsed between function invocation and having the result available in host memory. For the CUDA implementation this includes device memory allocation/ deallocation, memory transfer between device and host, and execution of the algorithm's main loop (i.e., the entire process of achieving the result). Each runtime measurement was gathered by running 10 full cycles of the algorithm.

For the evaluated (relatively small) problem sizes, the end-to-end runtime of the CUDA implementation varies by a very small amount, and is dominated not by the algorithm execution, but by the overhead from the memory management, data movement, and kernel launches. This is certainly not the case for the MATLAB program, whose end-to-end runtime increases drastically with an increase of the dimension size. 
For 3D tensors with dimension sizes $N=8, 16, 32$, the CUDA implementation proved to be approximately 2, 11, and 550 times faster (see Table \ref{tab:matlab_cuda_comparison}).

\begin{table}[t]
\centering
\caption{End-to-end execution time comparison between the MATLAB reference and the CUDA implementation for 3D tensors ($D=3$).}
\label{tab:matlab_cuda_comparison}
\begin{tabular}{crrr}
\hline
\textbf{Dimension Size ($N$)} & \textbf{MATLAB  (s)} & \textbf{CUDA (s)} & \textbf{Speedup} \\ \hline
8  & 0.64 & 0.36 & $1.8\times$     \\
16 & 4.21 & 0.37 & $11.3\times$    \\
32 & 214.45 & 0.39 & $551.3\times$   \\ \hline
\end{tabular}
\end{table}

\section{Conclusion and future work}
\label{sec:conclusion}
We developed a CUDA implementation of the Jacobi-type algorithm for tensor decomposition for arbitrary dense tensors of order $D\geq 3$ on a single CUDA-enabled GPU, without assuming symmetry or sparsity. Numerical experiments on tensors of up to nine dimensions confirmed convergence to the expected diagonal form and demonstrated runtime scaling on H100 GPUs as tensor order increases. A key finding is that the algorithm converges correctly even without pivot checks, despite these checks being the most computationally expensive part of the implementation, suggesting significant potential runtime reductions. 

As future work, we plan to extend the implementation to structured tensor cases, such as symmetric and skew-symmetric tensors. We also aim to demonstrate its applicability to spectral analysis of Heisenberg spin chains from Quantum Chemistry, including the observation of eigenvalue distributions.

\begin{credits}
\subsubsection{\ackname} 
This work was partially supported by eSSENCE, a Swedish strategic research program in e-Science. 
The computations were enabled by resources via the project UPPMAX 2025/2-247 provided by Uppsala University at UPPMAX.
\end{credits}

\bibliographystyle{splncs04}
\bibliography{references}

@article{Menczer25,
author = {Menczer, Andor and Legeza, {\"O}rs},
title = {Massively Parallel Tensor Network State Algorithms on Hybrid CPU-GPU Based Architectures},
journal = {J. Chem. Theory Comput.},
volume = {21},
number = {4},
pages = {1572-1587},
year = {2025},
doi = {10.1021/acs.jctc.4c00661}
}

@article{Borrelli_2020,
year = {2020},
publisher = {IOP Publishing},
volume = {22},
number = {12},
pages = {123002},
author = {Borrelli, Raffaele and Gelin, Maxim F},
title = {Quantum dynamics of vibrational energy flow in oscillator chains driven by anharmonic interactions},
journal = {New J. Phys.},
doi = {10.1088/1367-2630/abc9ed},
}

@article{RevModPhys,
  title = {The density-matrix renormalization group},
  author = {Schollw\"ock, U.},
  journal = {Rev. Mod. Phys.},
  volume = {77},
  issue = {1},
  pages = {259-315},
  numpages = {0},
  year = {2005},
  doi = {10.1103/RevModPhys.77.259},
}

@article{Larsson24,
author = {Henrik R. Larsson},
title = {A tensor network view of multilayer multiconfiguration time-dependent Hartree methods},
journal = {Mol. Phys.},
volume = {122},
number = {14},
pages = {e2306881},
year = {2024},
doi ={10.1080/00268976.2024.2306881},
}

@article{digitalquamtumspin26,
  title = {Digital Quantum Simulation of Spin Transport},
  author = {Lee, Yi-Ting and Pokharel, Bibek and Cohn, Jeffrey and Schleife, Andr\'e and Banerjee, Arnab},
  journal = {Phys. Rev. Lett.},
  volume = {136},
  issue = {5},
  pages = {050603},
  numpages = {7},
  year = {2026},
  doi = {10.1103/mx9k-kdlj},
}

@article{kolda-review09,
author = {Kolda, Tamara G. and Bader, Brett W.},
title = {Tensor Decompositions and Applications},
journal = {SIAM Review},
volume = {51},
number = {3},
pages = {455-500},
year = {2009},
doi = {10.1137/07070111X},
}

@misc{cutensor,
  author       = {NVIDIA Corporation},
  title        = {cuTENSOR Library},
  year         = {n.d.},
  howpublished = {\url{https://developer.nvidia.com/cutensor}},
  note         = {Accessed: 2026-05-06}
}

@article{begovic_kovac_convergence_2023,
	title = {Convergence of a {Jacobi}-type method for the approximate orthogonal tensor diagonalization},
	volume = {60},
	number = {1},
	journal = {Calcolo},
	author = {Begović Kovač, Erna},
	year = {2023},
	pages = {3},
    doi = {10.1007/s10092-022-00498-x}
}

@misc{nvidia_cuda_toolkit,
  author       = {NVIDIA},
  title        = {CUDA Toolkit Documentation},
  year         = {2025},
  howpublished = {\url{https://docs.nvidia.com/cuda/}},
  note         = {Accessed: 2025-12-10}
}

@article{Lathauwer00,
author = {De Lathauwer, Lieven and De Moor, Bart and Vandewalle, Joos},
title = {On the Best Rank-1 and Rank-(R1 ,R2 ,. . .,RN) Approximation of Higher-Order Tensors},
journal = {SIAM J. Matrix Anal. Appl.},
volume = {21},
number = {4},
pages = {1324-1342},
year = {2000},
doi = {10.1137/S0895479898346995},
}

@article{Lathauwer00-2,
author = {De Lathauwer, Lieven and De Moor, Bart and Vandewalle, Joos},
title = {A Multilinear Singular Value Decomposition},
journal = {SIAM J. Matrix Anal. Appl.},
volume = {21},
number = {4},
pages = {1253-1278},
year = {2000},
doi = {10.1137/S0895479896305696}
}

@article {DrmacVeselic07,
    AUTHOR = {Drma\v{c}, Zlatko and Veseli\'c, Kre\v{s}imir},
     TITLE = {New fast and accurate {J}acobi {SVD} algorithm. {I}},
   JOURNAL = {SIAM J. Matrix Anal. Appl.},
  FJOURNAL = {SIAM Journal on Matrix Analysis and Applications},
    VOLUME = {29},
      YEAR = {2007},
    NUMBER = {4},
     PAGES = {1322--1342},
      ISSN = {0895-4798,1095-7162},
   MRCLASS = {65F20 (15A12 15A18 15A23)},
  MRNUMBER = {2369298},
MRREVIEWER = {Dario\ Fasino},
       DOI = {10.1137/050639193},
}

@article {Novakovic14,
    AUTHOR = {Novakovi\'c, Vedran},
     TITLE = {A hierarchically blocked {J}acobi {SVD} algorithm for single
              and multiple graphics processing units},
   JOURNAL = {SIAM J. Sci. Comput.},
  FJOURNAL = {SIAM Journal on Scientific Computing},
    VOLUME = {37},
      YEAR = {2015},
    NUMBER = {1},
     PAGES = {C1--C30},
      ISSN = {1064-8275,1095-7197},
   MRCLASS = {65F15 (65Y05 65Y10)},
  MRNUMBER = {3301312},
MRREVIEWER = {Fatemeh\ Panjeh Ali Beik},
       DOI = {10.1137/140952429},
}

@article {LukPark89,
    AUTHOR = {Luk, Franklin T. and Park, Haesun},
     TITLE = {On parallel {J}acobi orderings},
   JOURNAL = {SIAM J. Sci. Statist. Comput.},
  FJOURNAL = {Society for Industrial and Applied Mathematics. Journal on
              Scientific and Statistical Computing},
    VOLUME = {10},
      YEAR = {1989},
    NUMBER = {1},
     PAGES = {18--26},
      ISSN = {0196-5204},
   MRCLASS = {65F30},
  MRNUMBER = {976158},
       DOI = {10.1137/0910002},
}

@article {ULC183,
    AUTHOR = {Li, Jianze and Usevich, Konstantin and Comon, Pierre},
     TITLE = {Globally convergent {J}acobi-type algorithms for simultaneous
              orthogonal symmetric tensor diagonalization},
   JOURNAL = {SIAM J. Matrix Anal. Appl.},
  FJOURNAL = {SIAM Journal on Matrix Analysis and Applications},
    VOLUME = {39},
      YEAR = {2018},
    NUMBER = {1},
     PAGES = {1--22},
      ISSN = {0895-4798,1095-7162},
   MRCLASS = {65F15 (15A69 49M30 90C30)},
  MRNUMBER = {3743743},
MRREVIEWER = {Zvi\ Retchkiman K\"onigsberg},
       DOI = {10.1137/17M1116295},
}

@article {ULC19,
    AUTHOR = {Li, Jianze and Usevich, Konstantin and Comon, Pierre},
     TITLE = {On approximate diagonalization of third order symmetric
              tensors by orthogonal transformations},
   JOURNAL = {Linear Algebra Appl.},
  FJOURNAL = {Linear Algebra and its Applications},
    VOLUME = {576},
      YEAR = {2019},
     PAGES = {324--351},
      ISSN = {0024-3795,1873-1856},
   MRCLASS = {15A69 (65F15 90C30 90C90)},
  MRNUMBER = {3958151},
MRREVIEWER = {Zvi\ Retchkiman K\"onigsberg},
       DOI = {10.1016/j.laa.2019.03.006},
}

@article {ULC20,
    AUTHOR = {Usevich, Konstantin and Li, Jianze and Comon, Pierre},
     TITLE = {Approximate matrix and tensor diagonalization by unitary
              transformations: convergence of {J}acobi-type algorithms},
   JOURNAL = {SIAM J. Optim.},
  FJOURNAL = {SIAM Journal on Optimization},
    VOLUME = {30},
      YEAR = {2020},
    NUMBER = {4},
     PAGES = {2998--3028},
      ISSN = {1052-6234,1095-7189},
   MRCLASS = {90C30 (15A69 53B20 53B21 65F15 65K10 65Y20)},
  MRNUMBER = {4164076},
       DOI = {10.1137/19M125950X},
}

@book {GolubVanLoan_book,
    AUTHOR = {Golub, Gene H. and Van Loan, Charles F.},
     TITLE = {Matrix computations},
   EDITION = {4th},
 PUBLISHER = {JHU Press},
      YEAR = {2013},
     PAGES = {xiv+756},
      ISBN = {978-1-4214-0794-4; 1-4214-0794-9; 978-1-4214-0859-0},
   MRCLASS = {65-02 (65Fxx)},
  MRNUMBER = {3024913},
MRREVIEWER = {J\"org\ Liesen},
}

@article {BegovicPerkovic24,
    AUTHOR = {Begovi\'c{} Kova\v{c}, Erna and Perkovi\'c, Ana},
     TITLE = {Trace maximization algorithm for the approximate tensor
              diagonalization},
   JOURNAL = {Linear Multilinear Algebra},
    VOLUME = {72},
      YEAR = {2024},
    NUMBER = {3},
     PAGES = {429--450},
      ISSN = {0308-1087,1563-5139},
   MRCLASS = {65F10 (15A15 15A69)},
  MRNUMBER = {4703579},
       DOI = {10.1080/03081087.2022.2158997},
}

\end{document}